\documentclass[12pt,twocolumn]{aastex63}
\usepackage{bm}
\accepted{March 2, 2021}
\submitjournal{ApJ}

\shorttitle{Biermann battery driven by a return current}
\shortauthors{Yutaka Ohira}
\begin{document}
\title{The Biermann battery driven by a streaming plasma}
\author{Yutaka Ohira}
\affiliation{Department of Earth and Planetary Science, The University of Tokyo, \\
7-3-1 Hongo, Bunkyo-ku, Tokyo 113-0033, Japan}
\email{y.ohira@eps.s.u-tokyo.ac.jp}

\begin{abstract}
The Biermann battery is a promising generation mechanism of a seed magnetic field in the universe. 
We propose a new driving mechanism for the Biermann battery in an inhomogeneous plasma with a beam component. 
An inhomogeneous electron flow induced by the return current can make an electron pressure gradient which is not parallel to the density gradient. 
As a result, the magnetic field is generated by the Biermann battery. 
We demonstrate the new generation mechanism of the magnetic field by conducting three-fluid plasma simulations. 
By this mechanism, the first cosmic rays can generate the seed magnetic field with astrophysical scales at redshift around $20$. 
\end{abstract}

\section{Introduction}
\label{sec:1}
Since the magnetic field plays various roles in laboratory, space, and astrophysical plasmas, the generation of the magnetic field has been widely discussed for a long time \citep{widrow02,subramanian19}. 
Nevertheless, it has not been well understood when, where, and how the magnetic field was first generated in the universe. 
The Biermann battery \citep{biermann50} is one of the promising mechanism of the magnetic field generation and has been widely investigated in many astrophysical phenomena \citep{kulsrud97,hanayama05,doi11,shiromoto14}.  and laboratory plasmas \citep{stamper91,gregori12}. 
In the Biermann battery mechanism, the electron pressure gradient which is not parallel to the density gradient makes a vortex flow of electrons, so that the electric current and the magnetic field are generated. 
To generate such a pressure structure, a distorted shock front or some electron heatings have been considered in two component (electrons and ions) plasmas \citep{subramanian94,kulsrud97,hanayama05,doi11,shiromoto14}. 
In this work, we propose a new driving mechanism for the Biermann battery. 

Plasmas sometimes have a streaming (or beam) component of charged particles in addition to an electron-ion plasma. 
The streaming plasma initially disturbs the charge and current neutrality conditions, making an electric field. 
The electric field accelerates the background electrons and makes a return current to neutralize the total change and the total current. 
If there is inhomogeneity in the system, the electron flow induced by the return current becomes nonuniform. 
Recently, we showed that the ram pressure gradient of the electron flow can generate the magnetic field like the Biermann battery \citep{ohira20}. 
In addition, we discussed a possibility that the streaming of the first cosmic rays can generate the seed of the astrophysical magnetic field by the new mechanism \citep{ohira20}. 
In that paper, we considered situations where the thermal pressure of electrons does not play an important role. 
In this work, we show that the electron flow induced by the return current makes a pressure structure that can drive the Biermann battery, so that the magnetic field is generated. 
Therefore, for the magnetic field generation in plasmas with a beam component, the electron flow and dynamics play an important role. 
This is a remarkable difference from work on the magnetic field generation so far. 

\section{Theory}
\label{sec:2}
We consider an electron-proton plasma with a beam plasma in a collisionless system. 
From the generalized Ohm's law for a multi-component plasma \citep{ohira20}, 
the electric field that generates magnetic field with a length scale larger than the electron inertial length scale is given by 
\begin{equation}
{\bm E} = - {\bm V}_{\rm e}  \times {\bm B} + \frac{m_{\rm e}}{e^2n_{\rm e}}{\bm \nabla} \cdot \left(\sum_s q_s n_s {\bm V}_s {\bm V}_s \right) - \frac{{\bm \nabla} p_{\rm e}}{en_{\rm e}}, 
\label{eq:ohm}
\end{equation}
%
where ${\bm E}, {\bm B}, q_s, n_s, {\bm V}_s, p_s$, and $m_s$ are the electric field, magnetic field, charge, number density, velocity, pressure, and mass, respectively. 
The subscript $s$ denotes the particle species.   
We use $s=$ e, p, and b for electrons, protons, and beam component in this work.
The first term represents advection of magnetic field by the electron flow, that is initially negligible as long as the magnetic field is zero or very weak. 
The second term describes the ram pressure and advection of each current \citep{ohira20}. 
The third term drives the Biermann battery effect.
The time evolution of the magnetic field is obtained from the Faraday's equation, $\partial {\bm B}/\partial t = - c{\bm \nabla}\times {\bm E}$, where $c$ is the speed of light. 

The time evolution of the electron pressure is described by the energy equation of fluid dynamics, 
\begin{equation}
\frac{\partial p_{\rm e}}{\partial t} + {\bm V}_{\rm e} \cdot {\bm \nabla}p_{\rm e}= - \gamma p_{\rm e} {\bm \nabla} \cdot {\bm V}_{\rm e},
\label{eq:pe}
\end{equation}
where $\gamma$ is the adiabatic index. 
The electron flow changes the initial distribution of the electron pressure, 
but does not change significantly the electron-density profile because the electron distribution has to be almost the same as that of ions to satisfy the charge neutrality condition. 
Therefore, even though the Biermann battery does not work initially ($\nabla p_{\rm e}\times \nabla n_{\rm e}=0$), the electron return current can make a pressure structure in which the Biermann battery works ($\nabla p_{\rm e}\times \nabla n_{\rm e}\neq 0$). 
This driving mechanism for the Biermann battery has not been considered so far. 

To describe the new magnetic field generation in detail, we specify the plasma condition that we consider in this work. 
The initial magnetic field is zero.
The plasma consists of electrons, protons, and beam ions with the charge of $q_{\rm b} =Z_{{\rm b}}e$.
The protons and electrons have a nonuniform density distribution that is organized by entropy modes. 
Namely, there is initially no pressure gradient. 
All quantities of the beam ions are uniform. 
The beam direction is set to be the x-axis. 
Then, in the proton rest frame, the charge and current neutrality conditions become
\begin{equation}
n_{\rm e} = n_{\rm p}+Z_{\rm b}n_{\rm b}, \ \ n_{\rm e}{\bm V}_{\rm e}=Z_{\rm b}n_{\rm b} V_{\rm b}{\bm e}_{\rm x} ,
\end{equation}
where ${\bm e}_{\rm x}$ is the unit vector in the x direction.
It should be noted that the current of beam ions is uniform but the electron velocity field, ${\bm V}_{\rm e}$, is not uniform. 
By assuming that the electron flow is constant in time and the pressure gradient is small to neglect the second term on the left hand side of Equation~(\ref{eq:pe}) safely, the time evolution of the electron pressure is approximately given by 
\begin{equation} 
p_{\rm e}= p_{\rm e,0} \exp \left( - \gamma t \frac{\partial V_{\rm e}}{\partial x} \right) , 
\label{eq:pe2}
\end{equation}
where $p_{\rm e,0}$ is the initial electron pressure. 
Then, the time evolution of the magnetic field is given by 
\begin{equation}
\frac{\partial {\bm B}}{\partial t} =  \frac{m_{\rm e}c}{2e} {\bm \nabla} \times \frac{\partial V_{\rm e}^2}{\partial x} {\bm e}_{\rm x} - \frac{cp_{\rm e}\gamma t}{en_{\rm e}V_{\rm e}} {\bm \nabla}V_{\rm e} \times {\bm \nabla}\frac{\partial V_{\rm e}}{\partial x}.
\label{eq:mag}
\end{equation}
The first term describes the magnetic field generation by the ram pressure gradient of the electron flow \citep{ohira20}.
The magnetic field generation by the second term describes what we newly propose in this work, which is derived from the pressure gradient term on the right hand side of Equation~(\ref{eq:ohm}) and Equation~(\ref{eq:pe2}). 
As long as $\gamma t\partial V_{\rm e}/\partial x \ll 1$, the first and second terms can be considered to be constant in time and proportional to time, respectively. 
Therefore, the new mechanism proposed in this work eventually dominates the magnetic field generation. 

As a simple example, we consider the following initial condition: 
\begin{eqnarray}
n_{\rm e} &=& n_{\rm e,0} \left [1+\delta \left\{ \sin \left( \frac{2\pi}{L}x \right) +\sin \left( \frac{2\pi}{L}y \right) \right \} \right ]^{-1} , \nonumber \\ \
n_{\rm p} &=& n_{\rm e}  - Z_{\rm b}n_{\rm b,0} ,\nonumber \\ \
n_{\rm b} &=& n_{\rm b,0},\nonumber \\ \
p_{\rm e} &=& p_{\rm p} = p_{\rm e,0},\nonumber \\ \
p_{\rm b} &=& p_{\rm b,0},\nonumber \\ \
{\bm V}_{\rm e} &=& V_{\rm e,0} \left (\frac{n_{\rm e,0}}{n_{\rm e}} \right) {\bm e}_{\rm x} ,\nonumber \\ \
{\bm V}_{\rm p} &=& {\bm 0},\nonumber \\ \
{\bm V}_{\rm b} &=& V_{\rm e,0} \left(\frac{n_{\rm e,0}}{Z_{\rm b}n_{\rm b,0}}\right){\bm e}_{\rm x}, \nonumber 
\end{eqnarray}
where the charge and current neutrality conditions have been satisfied and 
$n_{\rm e,0}, n_{\rm b,0}, p_{\rm e,0}, p_{\rm b,0}, V_{\rm e,0}, \delta$, and $L$ are constants. 
We impose the periodic boundary condition in both x and y directions. 
Since ${\bm V}_{\rm e}$ and $p_{\rm e}$ are safely assumed to be constant in time for $t\ll L/V_{\rm e,0}$, the analytical solution to Equation~(\ref{eq:mag}) is given by
\begin{eqnarray}
{\bm B} &=&  -\frac{4\pi^2m_{\rm e}cV_{\rm e,0}\delta^2}{eL} \cos \left(\frac{2\pi}{L}y\right) {\bm e}_{\rm z}  \nonumber \\ \
&\times&  \left\{ \left(\frac{V_{\rm e,0}t}{L}\right)  \cos \left(\frac{2\pi}{L}x\right)+ \pi \left(\frac{c_{\rm se}t}{L}\right)^2\sin\left(\frac{2\pi}{L}x\right)  \right \} ,
\label{eq:ana1}
\end{eqnarray}
where $c_{\rm se}=(\gamma p_{\rm e,0}/m_{\rm e}n_{\rm e,0})^{1/2}$ is the sound velocity of electrons. 
The magnetic field is initially generated by the ram pressure gradient \citep{ohira20}, and the generation rate depends on the electron flow velocity. 
After $t = LV_{\rm e,0}/\pi c_{\rm se}^2$, the Biermann battery driven by the electron flow mainly generates the magnetic field. 
In the later phase, the magnetic field increases with the square of time and the generation rate depends on the electron temperature (or electron sound velocity) and the electron flow velocity. 
The time evolution of the spatially averaged magnetic field energy density becomes  
\begin{eqnarray}
\frac{<B_z^2>}{4\pi n_{\rm e,0}m_{\rm e}V_{\rm e,0}^2} &=& 4\pi^4\delta^4 \left(\frac{c}{L\omega_{\rm pe}} \right)^{2} \nonumber \\ \
&\times&\left \{ \left(\frac{V_{\rm e,0}t}{L}\right)^2+\pi^2 \left(\frac{c_{\rm se}t}{L}\right)^4 \right\}, 
\label{eq:ana2}
\end{eqnarray}
where we normalized the magnetic field energy density by the kinetic energy density of the electron drift motion. $\omega_{\rm pe}$ is the electron plasma frequency.

\section{Simulation}
\label{sec:3}
\begin{figure}
\begin{center}
\epsscale{1.2}
\plotone{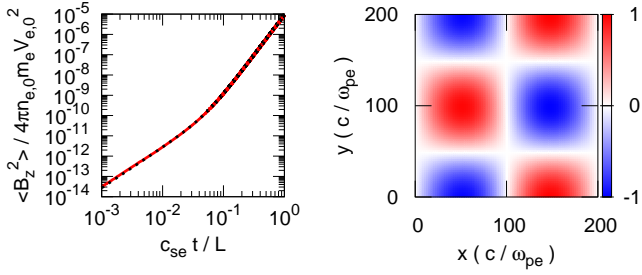}
\caption{{\it Left}: time evolution of the spatially averaged energy density of magnetic fields. The red solid and black dashed lines show the simulation result and the analytical solution, Equation~(\ref{eq:ana2}), respectively. {\it Right}: simulation result of the z component of magnetic fields at $t=0.3\ L/c_{\rm se}$. The magnetic fields are normalized by the maximum value of the analytical solution, Equation~(\ref{eq:ana1}).}
\label{f1}
\end{center}
\end{figure}
\begin{figure}
\begin{center}
\epsscale{1.1}
\plotone{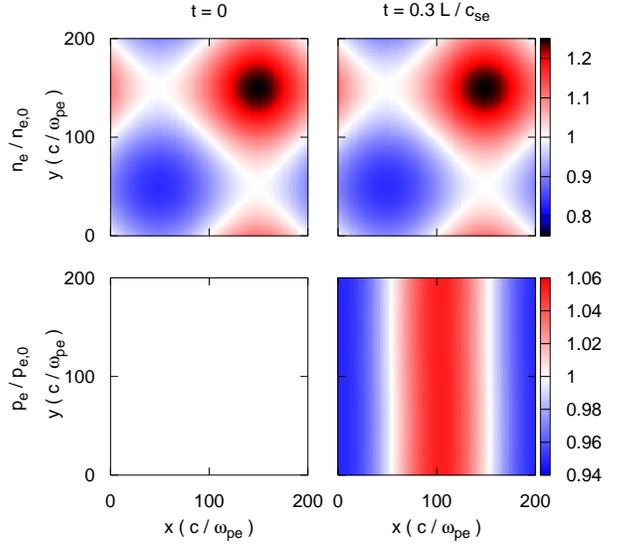}
\caption{Simulation results of the electron density (top two panels) and electron pressure (bottom two panels) at $t=0$ (left column) and $t=0.3\ L/c_{\rm se}$ (right column).}
\label{f2}
\end{center}
\end{figure}
In oder to verify the magnetic field generation described by Equation~(\ref{eq:mag}), we perform a two-dimensional multi-fluid simulation which solves the full Maxwell equations and fluid equations for each species \citep{hakim06}. 
The generalized Ohm's law is not assumed in this simulation. 
The simulation parameters are as follows: $n_{\rm b,0}/n_{\rm e,0}=2\times10^{-3}, V_{\rm b,0}=0.25\ c, V_{\rm e,0}=5\times 10^{-4}\ c, \delta=0.1, L=200\ c/\omega_{\rm pe}, m_{\rm p}/m_{\rm e}=m_{\rm b}/m_{\rm e}=1836, Z_{\rm b}=1, c_{\rm se}=3\times10^{-3}c, c_{\rm sp}=c_{\rm se}(m_{\rm e}/m_{\rm p})^{1/2}, c_{\rm sb}=0.1c$, and $\gamma_{\rm e}=\gamma_{\rm p}=\gamma_{\rm b}=5/3$. 
The cell size and the time step are set to $\Delta x = 0.5 \ c/\omega_{\rm pe}$ and $\Delta t = 0.1\ \omega_{\rm pe}^{-1}$, respectively.  
Although the simulation box size is much smaller than the astrophysical scale but larger than the proton inertial length scale, 
we can scale the results in the small simulation box to results in the astrophysical scale.
This is because there is no characteristic length scale over the proton inertial length scale, $(m_{\rm p}/m_{\rm e})^{1/2}c/\omega_{\rm pe}\approx 43\ c/\omega_{\rm pe}$, in collisionless unmagnetized plasma.

The left panel of Figure~\ref{f1} shows the time evolution of the spatially averaged energy density of magnetic fields. 
The red solid line shows the simulation result for the z component of magnetic fields, and the black dashed line shows the analytical solution, Equation~(\ref{eq:ana2}). 
The right panel of Figure~\ref{f1} shows the simulation result of the z component of magnetic fields at time $t=0.3\ L/c_{\rm se}$.  
The magnetic fields are normalized by the maximum value of the analytical solution, Equation~(\ref{eq:ana1}). 
All the simulation results are excellently in agreement with the analytical solutions, Equations~(\ref{eq:ana1}) and (\ref{eq:ana2}). 
Note that the value of the vertical axis in the left panel of Figure~\ref{f1} has to be scaled when we apply the simulation result to astrophysical phenomena. 
Since the magnetic field energy fraction is proportional to $(c/L\omega_{\rm pe})^2$ (see Equation~(\ref{eq:ana2})), 
it becomes at least $10^{-22}$ times smaller than the simulation result if $L$ is kpc scale. 
In addition, this simulation assumed unrealistic value of $V_{\rm e,0}/c_{\rm se}\approx 0.17$. 
For a more realistic value of $V_{\rm e,0}/c_{\rm se}\sim 5\times 10^{-4}$ (for details see Section~\ref{sec:4}), 
the contribution of the Biermann battery by the ram pressure (first term in Equation (\ref{eq:ana2})) is negligible for $c_{\rm se}t/L >10^{-4}$.

Figure~\ref{f2} shows distributions of the electron density (top two panels) and the electron pressure (bottom two panels) at $t=0$ (left column) and $t=0.3\ L/c_{\rm se}$ (right column). 
The electron density does not change from the initial state during this simulation because heavy protons almost do not move during the simulation time scale 
and electrons seek to satisfy the charge neutrality condition. 
However, the electron pressure distribution changes as time goes on because the electron flow has compressive modes. 
The electron pressure is uniform in the initial condition (left bottom panel). 
Then, a nonuniform structure is generated (right bottom panel), $p_{\rm e}/p_{\rm e,0}-1\approx -\gamma t (\partial V_{\rm e}/\partial x)\propto -\cos(2\pi x/L)$ (see Equation~(\ref{eq:pe2})). 
As a result, the Biermann battery term ($\propto \nabla p_{\rm e}\times \nabla n_{\rm e}$) increases with time and makes the magnetic field in the later phase. 
Hence, we confirmed that the electron return current drives the Biermann battery mechanism in addition to the magnetic field generation by the ram pressure of the electron flow. 

\section{Discussion}
\label{sec:4}
We discuss the saturation of the magnetic field generation. 
Since the generated magnetic field is advected by the electron flow in this mechanism, 
the magnetic field generation can work during the advection time scale, $L/V_{\rm e}$. 
In this case, the ratio of the gyroradius of thermal protons, $r_{\rm gp}=cm_{\rm p}c_{\rm sp}/eB$ to the coherent length scale of the magnetic field, $L$, is represented by 
\begin{equation}
\frac{r_{\rm gp}}{L}= \frac{1}{4\pi^3}\left(\frac{V_{\rm e,0}}{c_{\rm sp}}\right)~~,
\label{eq:magnetization}
\end{equation}
where $c_{\rm cp}$ is the sound (or thermal) velocity of protons. 
Since the ratio of $V_{\rm e,0}/c_{\rm sp}$ is usually less than unity, 
the plasma is strongly magnetized ($r_{\rm gp}/L\ll 1$) at the end of the magnetic field generation. 
Therefore, the generated magnetic field can be further amplified by some magneto-hydrodynamical dynamo 
after the generation of the magnetic field \citep{rincon16,pusztai20}.
It should be noted that the magnetic field generation is sometimes limited by a finite time smaller than $L/V_{\rm e,0}$. 
Even in such a case, the plasma can be magnetized if the condition of $r_{\rm gp}/L < 1$ is satisfied.

In our previous work \citep{ohira20}, we estimated the magnetic field strength generated by the first term in Equation~(\ref{eq:mag}) in the early universe. 
Since we found that the second term in Equation~(\ref{eq:mag}) becomes more important for a later phase, we here re-estimate it.  
The first cosmic rays are accelerated by the first supernova remnant at redshift $z \approx 20$ \citep{ohira19}.
After the acceleration, the first cosmic rays propagate to the intergalactic medium with almost the speed of light. 
The propagating first cosmic rays induce the electron return current whose velocity is estimated to be $V_{\rm e}\sim10^4 {\rm cm\ s}^{-1}$ at $z\approx 20$ \citep{ohira20}. 
From Equation~(\ref{eq:ana1}), the magnetic field strength at $z\approx 20$ generated by the new mechanism proposed in this work is 
\begin{eqnarray}
B\sim 5.5\times 10^{-21}\ &{\rm G}& \left(\frac{\delta}{1}\right)\left(\frac{V_{\rm e,0}}{10^4\ {\rm cm\ s}^{-1}}\right) \left(\frac{T_{\rm e,0}}{0.1\ {\rm eV}}\right)  \nonumber \\ 
&\times&   \left(\frac{L}{1\ {\rm kpc}}\right)^{-3}  \left(\frac{t}{10^8\ {\rm yr}}\right)^{2} ~~,
\label{eq:bo}
\end{eqnarray}
where $T_{\rm e,0}$ is the electron temperature. 
This is three orders of magnitude larger than estimated by our previous work \citep{ohira20} and sufficiently large as the seed of the magnetic field in the current galaxies \citep{davis99}. 
For the above parameters, the magnetic field generation is limited not by the advection time scale ($L/V_{\rm e,0}$) but by a finite time scale of $10^8$ yr. 
Although Equation~(\ref{eq:magnetization}) cannot be satisfied and $r_{\rm gp}/L \approx 2.8$, the magnetization of $r_{\rm gp}/L < 1$ can be satisfied 
for $L=0.5\ {\rm kpc}$ or $2\times 10^8\ {\rm yr}$. 

\begin{figure}
\begin{center}
\epsscale{1.1}
\plotone{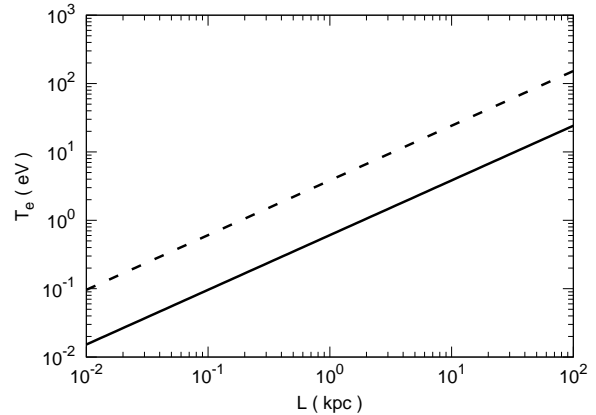}
\caption{Parameter space of the electron temperature ($T_{\rm e}$) and size ($L$) for determining the relative importance between our mechanism and the resistive mechanism. Our mechanism dominates the resistive mechanism in the region above each the line. The solid and dashed lines are for $t=10^8\ {\rm yr}, n_{\rm e}=10^{-7}\ {\rm cm}^{-3}$ and for $t=10^9\ {\rm yr}, n_{\rm e}=10^{-4}\ {\rm cm}^{-3}$, respectively. }
\label{f3}
\end{center}
\end{figure}

There is another mechanism of the magnetic field generation in plasmas with a beam component. 
\citet{bell03} showed that the magnetic field can be generated if there is a non-uniform resistivity between background electrons and ions. 
The Coulomb collision between the background electrons and ions reduces the return current but the current of the beam plasma does not change 
because the beam plasma has a high velocity. 
As a result, the return current cannot cancel out the the beam current and the magnetic field is generated by the total current. 
\citet{miniati11} applied the resistive magnetic field generation to the intergalactic space by considering cosmic rays as the beam plasma. 
For this mechanism, the Ohm's law is ${\bm E}=\eta en_{\rm b}V_{\rm b}{\bm e}_{\rm x}$ in our formalism, where $\eta\approx 3.7\times 10^{-12}(T_{\rm e}/0.1{\rm eV})^{-3/2}\ {\rm sec}$ is the Spitzer resistivity. If the electron temperature has inhomogeneity, $T_{\rm e}=T_{\rm e,0}(1+\delta)\sin(2\pi y/L)$, the magnetic field strength can be estimated by
\begin{eqnarray}
B\sim 5.0\times 10^{-19}\ &{\rm G}& \left(\frac{\delta}{1}\right)\left(\frac{n_{\rm b}V_{\rm b}}{10^{-3}\ {\rm cm^2\ s}^{-1}}\right) \left(\frac{T_{\rm e,0}}{0.1\ {\rm eV}}\right)^{-3/2}  \nonumber \\ 
&\times&   \left(\frac{L}{1\ {\rm kpc}}\right)^{-1}  \left(\frac{t}{10^8\ {\rm yr}}\right) ~~.
\label{eq:bb}
\end{eqnarray}
Compareing Equations (\ref{eq:bo}) and (\ref{eq:bb}), one can obtain the condition in order for our mechanism to dominate the resistive magnetic field generation, 
\begin{equation}
T_{\rm e,0} > 0.6\ {\rm eV} \left(\frac{L}{1\ {\rm kpc}}\right)^{4/5} \left(\frac{n_{\rm e}}{10^{-7}\ {\rm cm}^{-3}}\right)^{2/5} \left(\frac{t}{10^8\ {\rm yr}}\right)^{2/5}.
\end{equation}
This condition is plotted in Figure~\ref{f3}.  
The solid line is for $t=10^8\ {\rm yr}, n_{\rm e}=10^{-7}\ {\rm cm}^{-3}$, which corresponds to the gas in the Universe before the reionization era. 
The dashed line is for $t=10^9\ {\rm yr}, n_{\rm e}=10^{-4}\ {\rm cm}^{-3}$, which corresponds to that after the reionization era. 
Our mechanism dominates the resistive mechanism in the region above each the line.
Basically, our mechanism efficiently generates the magnetic field in hot and small scale regions, while the resistive magnetic field generation efficiently works in cold and large scale regions. 
Therefore, both mechanisms work in different region of the Universe. 
Comprehensive studies about our mechanism should be done in the framework of the cosmological structure formation 
as \citet{miniati11} investigated the resistive mechanism.

There are many beam plasmas in current astrophysical plasmas. 
In particular, collisionless shocks make the beam plasmas by several ways, reflection of upstream plasmas \citep{leroy83}, leaking of downstream plasmas \citep{spitkovsky08,tomita19}, and the charge exchange process \citep{ohira09b,ohira12,blasi12,ohira13}.
In addition, accelerating and escaping cosmic rays can be regarded as the beam plasma \citep{axford77,krymsky77,bell78,blandford78,ohira10,fujita10,ohiraetal12}.
These beam plasmas excite many kinetic plasma instabilities around the collisionless shocks. 
The magnetic field amplification by the kinetic instabilities are actively studied to understand the cosmic-ray acceleration \citep{bell04,reville08,niemiec08,riquelme09,ohira09a}. 
For acceleration of protons and ions, hybrid-type simulations are widely performed, where electrons are treated as a massless fluid but ions or nonthermal ions are treated as particles \citep{caprioli16,ohira16a,ohira16b,bai15,vanmarle18,vanmarle19}. 
Since the massless-electron model cannot solve dynamics of electrons correctly, 
those hybrid-type simulations cannot solve the magnetic field generations proposed in this work and our previous work \citep{ohira20}. 
The generation mechanism of the magnetic field in this work could be important for the cosmic-ray acceleration at collisionless shocks.

\section{Summary}
\label{sec:5}
We have presented a new driving mechanism for the Biermann battery which generates magnetic field. 
If a beam plasma is propagating to an inhomogeneous plasma, 
the electron flow induced by the return current has compressible modes for some cases. 
Then, a gradient of the electron pressure is generated by the compressible flow of electrons, 
but the electron density structure is not advected by the compressible flow and does not change from the initial ion density structure because of the charge neutrality. 
As a result, the electron vortex is generated and the magnetic field is generated by the Biermann battery mechanism. 
By performing multi-fluid plasma simulations, we have confirmed the new driving mechanism of the Biermann battery. 
The first cosmic ray can generate sufficiently large magnetic field by this mechanism.  
The expected magnetic field strength at redshift around $20$ is $B\sim 5.5\times 10^{-21}\ {\rm G}$ in kpc scales.

\acknowledgments
We thank the referee for valuable comments to improve the paper. 
Numerical computations were carried out on the XC50 system at the Center for Computational Astrophysics (CfCA) of the National Astronomical Observatory of Japan. 
This work is supported by JSPS KAKENHI Grant Number JP16K17702 and JP19H01893, and by Leading Initiative for Excellent Young Researchers, MEXT, Japan.


\begin{thebibliography}{}

\bibitem[Axford et al.(1977)]{axford77}
Axford, W. I., Leer, E., \& Skadron, G., 1977, Proc. 15th Int. Cosmic Ray Conf., (Plovdiv: Bulgarian Academy of Sciences), 11, 132

\bibitem[Bai et al.(2015)]{bai15}
Bai, X.-N., Caprioli, D., Sironi, L., \&Spitkovsky, A., 2015, \apj, 809, 55

\bibitem[Bell(1978)]{bell78}
Bell, A. R., 1978, \mnras, 182, 147

\bibitem[Bell \& Kingham(2003)]{bell03}
Bell, A. R., \& Kingham, R. J., 2003, \prl, 91, 035003

\bibitem[Bell(2004)]{bell04}
Bell, A. R., 2004, \mnras, 353, 550

\bibitem[Biermann(1950)]{biermann50}
Biermann, L., 1950, Zeitschrift f\"{u}r Naturforschung A, 5, 65

\bibitem[Blasi et al.(2012)]{blasi12}
Blasi, P., Morlino, G., Bandiera, R., Amato, E., \& Caprioli, D., \apj, 755, 121

\bibitem[Blandford \& Ostriker(1978)]{blandford78}
Blandford, R. D., \& Ostriker. J. P., 1978, \apj, 221, 29

\bibitem[Caprioli \& Spitkovsky(2016)]{caprioli16}
Caprioli, D., \& Spitkovsky, A., 2016, \apjl, 817, 137

\bibitem[Davis et al.(1999)]{davis99}
Davis, A. C., Lilley, M. \& T\"{o}rnkvist, O., \prd, 60, 021301(R)

\bibitem[Doi \& Susa(2011)]{doi11}
Doi, K., \& Susa, H., 2011, \apj, 741, 93

\bibitem[Fujita et al.(2010)]{fujita10}
Fujita, Y., Ohira, Y., \& Takahara., F., 2010, \apjl, 712, L153

\bibitem[Gregori et al.(2012)]{gregori12}
Gregori, G., et al., 2012, \nat, 481, 480

\bibitem[Hakim et al.(2006)]{hakim06}
Hakim, A., Loverich, J., \& Shumlak, U., 2006, Journal of Computational Physics, 219, 418

\bibitem[Hanayama et al.(2005)]{hanayama05}
Hanayama, H., Takahashi, K., Kotake, K., Oguri, M., Ichiki, K., \& Ohno, H., 2005, \apj, 633, 93

\bibitem[Kulsrud et al.(1997)]{kulsrud97}
Kulsrud, R. M., Cen, R., Ostriker, J. P., \& Ryu, D., 1997, \apj, 480, 481

\bibitem[Krymsky (1977)]{krymsky77}
Krymsky, G. F., 1977, Dokl. Akad. Nauk SSSR, 234, 1306

\bibitem[Leroy(1983)]{leroy83}
Leroy, M. M., 1983, Phys. Fluids, 26, 2742

\bibitem[Miniati \& Bell(2011)]{miniati11}
Miniati, F., \& Bell, A. R., 2011, \apj 729, 73.

\bibitem[Niemiec et al.(2008)]{niemiec08}
Niemiec, J., Pohl, M., Stroman, T., Nishikawa, K., 2008, \apj, 684, 1174

\bibitem[Ohira et al.(2009a)]{ohira09a}
Ohira, Y., Reville, B., Kirk, J.G., \& Takahara, F., 2009a, \apj, 698, 445

\bibitem[Ohira et al.(2009b)]{ohira09b}
Ohira, Y., Takahara, F., \& Terasawa, T., 2009b, \apjl, 703, L59

\bibitem[Ohira et al.(2010)]{ohira10}
Ohira, Y., Murase, K., \& Yamazaki, R., 2010 \aap, 513, A17

\bibitem[Ohira et al.(2012)]{ohiraetal12}
Ohira, Y., Yamazaki, R., Kawanaka, N., \& Ioka, K., 2012 \mnras, 427, 91

\bibitem[Ohira(2012)]{ohira12}
Ohira, Y., 2012, \apj, 758, 979

\bibitem[Ohira(2013)]{ohira13}
Ohira, Y., 2013, \prl, 111, 245002

\bibitem[Ohira(2016a)]{ohira16a}
Ohira, Y. 2016 \apj, 817, 137

\bibitem[Ohira (2016b)]{ohira16b}
Ohira, Y. 2016, \apj, 827, 36

\bibitem[Ohira \& Murase(2019)]{ohira19}
Ohira, Y., \& Murase, K., 2019, \prd, 100, 061301(R)

\bibitem[Ohira(2020)]{ohira20}
Ohira, Y., 2020, \apjl, 896, L12

\bibitem[Pusztai et al.(2020)]{pusztai20}
Pusztai, I., Juno, J., Brandenburg, A., TenBarge, J. M., Hakim, A., Francisquez, M., \& Sundstr{\"o}n, A., 2020, \prl 124, 255102

\bibitem[Riquelme \& Spitkovsky(2009)]{riquelme09}
Riquelme, M. A. \& Spitkovsky, A., 2009, \apj, 694, 626

\bibitem[Reville et al.(2008)]{reville08}
Reville, B., O'Sullivan, S., Duffy, P., \& Kirk, J. G., 2008, \mnras 386, 509

\bibitem[Rincon et al.(2016)]{rincon16}
Rincon, F., Califano, F., Schekochihin, A. A., \& Valentini, F., 2016, PNAS, 113, 3950

\bibitem[Shiromoto et al.(2014)]{shiromoto14}
Shiromoto, Y., Susa, H., \& Hosokawa, T., 2014, \apj, 782, 108

\bibitem[Spitkovsky(2008)]{spitkovsky08}
Spitkovsky, A., 2008, \apjl, 673, L39

\bibitem[Stamper (1991)]{stamper91}
Stamper, J.A., 1991, Laser Part. Beams. 9, 841

\bibitem[Subramanian et al.(1994)]{subramanian94}
Subramanian, K., Narasimha, D., \& Chitre, S. M., 1994, \mnras 271, L15

\bibitem[Subramanian(2019)]{subramanian19}
Subramanian, K., 2019, Galaxies, 7, 47

\bibitem[Tomita et al.(2019)]{tomita19}
Tomita, S., Ohira, Y., \& Yamazaki, R., 2019, \apj, 886, 54

\bibitem[van Marle et al.(2018)]{vanmarle18}
van Marle, A. J., Casse, F., \& Marcowith, A., 2018, \mnras, 473, 3394

\bibitem[van Marle et al.(2019)]{vanmarle19}
van Marle, A. J., Casse, F., \& Marcowith, A., 2019, \mnras, 490, 1156

\bibitem[Widrow(2002)]{widrow02}
Widrow, L. M., 2002, Rev. Mod. Phys., 74, 775


\end{thebibliography}

\end{document}